\documentclass{article}

\usepackage{arxiv}
\usepackage[utf8]{inputenc}
\usepackage[T1]{fontenc}
\usepackage{url}
\usepackage{booktabs}
\usepackage{amsmath}
\usepackage{amsfonts}
\usepackage{microtype}
\usepackage{graphicx}
\usepackage{cite}
\usepackage{hyperref}

\graphicspath{{figures/}}
\newcommand{\githuburl}{https://github.com/douyipu/mosaic}

\title{MOSAIC: Multi-Objective Slice-Aware Iterative Curation for Alignment}

\author{
  \href{https://orcid.org/0009-0000-3126-1914}{\includegraphics[scale=0.06]{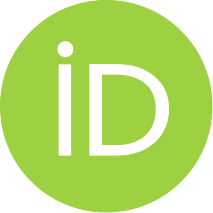}\hspace{1mm}Yipu~Dou}\thanks{Co-corresponding authors.}\\
  School of Cyber Science and Engineering\\
  Southeast University\\
  Nanjing 211189, China \\
  \texttt{scholar@douyipu.com} \\
  \And
  Wang~Yang\footnotemark[1] \\
  School of Cyber Science and Engineering\\
  Southeast University\\
  Nanjing 211189, China \\
  \texttt{wang.yang@seu.edu.cn} \\
}

\renewcommand{\shorttitle}{MOSAIC: Multi-Objective Slice-Aware Iterative Curation for Alignment}

\hypersetup{
  pdftitle={MOSAIC: Multi-Objective Slice-Aware Iterative Curation for Alignment},
  pdfsubject={cs.CL, cs.LG, cs.AI},
  pdfauthor={Yipu~Dou, Wang~Yang},
  pdfkeywords={supervised fine-tuning, data mixture search, safety alignment},
}

\begin{document}
\maketitle
\setlength{\headheight}{24pt}

\begin{abstract}
We study how to allocate a fixed supervised fine-tuning budget when three objectives must be balanced at once: multi-turn safety alignment, low over-refusal on benign boundary queries, and instruction following under verifiable constraints. We propose \textbf{MOSAIC} (\emph{Multi-Objective Slice-Aware Iterative Curation for Alignment}), a multi-objective framework for closed-loop data mixture search built on a unified L1--L3 evaluation interface. MOSAIC turns slice-level failure profiles into executable data actions, including dataset-level mixture ratios, bucket-level weights, and focus criteria. Under a fixed 1M-token budget and five rounds of independent fine-tuning from the same base model, MOSAIC improves internal XGuard from 2.76 to 4.67 while keeping OrBench at 4.41 and IFEval at 3.65. The final Pareto solution also generalizes better than a random static LoRA baseline on independent attack, over-refusal, and capability tests, suggesting that structured failure diagnosis can serve as a practical control signal for budgeted data construction. Code is available at \url{\githuburl}.
\end{abstract}


\section{Introduction}

Supervised fine-tuning (SFT) is increasingly constrained by data allocation rather than raw sample volume. Recent work such as LIMA, AlpaGasus, and Deita shows that, under fixed compute, relatively small but well-chosen data can outperform larger noisy collections \cite{DBLP:conf/nips/ZhouLX0SMMEYYZG23,DBLP:conf/iclr/ChenLYWGYTS0HJ24,DBLP:conf/iclr/Z010H24}. The open question is how to allocate limited budget when alignment requires several competing capabilities at once.

This trade-off is particularly sharp for models that must be safe, helpful, and instruction-faithful at the same time. Adding more safety data can reduce harmful compliance, but it can also increase over-refusal on benign edge cases or erode instruction following. Standard evaluation summaries are poorly suited to this setting: they can rank models, but they rarely tell us which slices failed or what data should be added next.

We address this problem with \textbf{MOSAIC} (\emph{Multi-Objective Slice-Aware Iterative Curation for Alignment}), a closed-loop data mixture search framework built around a unified L1--L3 annotation interface for XGuard, OrBench, and IFEval. The interface emits slice labels, importance weights, atomic failures, and aggregated scores, while a proposal agent converts the resulting failure profile into the next round's data actions under a fixed token budget. To isolate the effect of data distribution from training path dependence, every round restarts from the same base model and performs an independent LoRA fine-tuning run \cite{DBLP:conf/iclr/HuSWALWWC22}.

The main contributions are:
\begin{enumerate}
    \item We introduce a unified L1--L3 evaluation interface that makes safety alignment, over-refusal calibration, and instruction following comparable and diagnosable under a shared annotation language.
    \item We propose a closed-loop mapping from evaluation failures to executable data actions, including dataset-level mixture ratios, bucket-level weights, and focus criteria.
    \item We demonstrate that fixed-budget mixture search produces better balanced solutions than a random static LoRA baseline, both on internal search metrics and on independent external evaluations.
\end{enumerate}

\section{Related Work}

The first relevant line of work studies the construction and filtering of instruction-tuning data. InstructGPT, Self-Instruct, Alpaca, and later data-centric studies such as LIMA, AlpaGasus, and Deita all emphasize that alignment quality depends strongly on data composition rather than scale alone \cite{DBLP:conf/nips/Ouyang0JAWMZASR22,DBLP:conf/acl/WangKMLSKH23,alpaca,DBLP:conf/nips/ZhouLX0SMMEYYZG23,DBLP:conf/iclr/ChenLYWGYTS0HJ24,DBLP:conf/iclr/Z010H24}. Our work is aligned with this perspective, but optimizes the \emph{distribution} of a fixed training budget rather than the quality of isolated samples.

The second line of work focuses on automatic evaluation. MT-Bench, G-Eval, CheckEval, and IFEval all aim to make model evaluation more stable, interpretable, or programmatically verifiable \cite{DBLP:conf/nips/ZhengC00WZL0LXZ23,DBLP:conf/emnlp/LiuIXWXZ23,DBLP:conf/emnlp/LeeKKCKKK25,zhou2023instructionfollowingevaluationlargelanguage}. We build on this direction, but require the evaluation output to be actionable: scores must decompose into slice-level failure profiles that can drive the next round of data allocation.

The third line studies iterative or closed-loop data optimization. Recent systems such as LoopTool and Middo use model feedback to rewrite, filter, or resample data during training \cite{zhang2025looptool,tang2025middo}. Our setting differs in three ways. First, we optimize jointly over three alignment objectives rather than one. Second, we operate under a fixed token budget and fixed iteration count. Third, we restart every round from the same base model so that observed changes can be attributed to data distribution rather than inherited weights.

\section{The MOSAIC Framework}

\subsection{Problem Setting}

We treat the training set as a controllable distribution rather than a static corpus. The training pool is partitioned into slices derived from structured annotations, and each round chooses how much budget to allocate to each slice. The system is closed-loop: evaluation produces failure profiles, and those failure profiles determine the next training distribution.

\begin{figure}[t]
    \centering
    \includegraphics[width=0.98\linewidth]{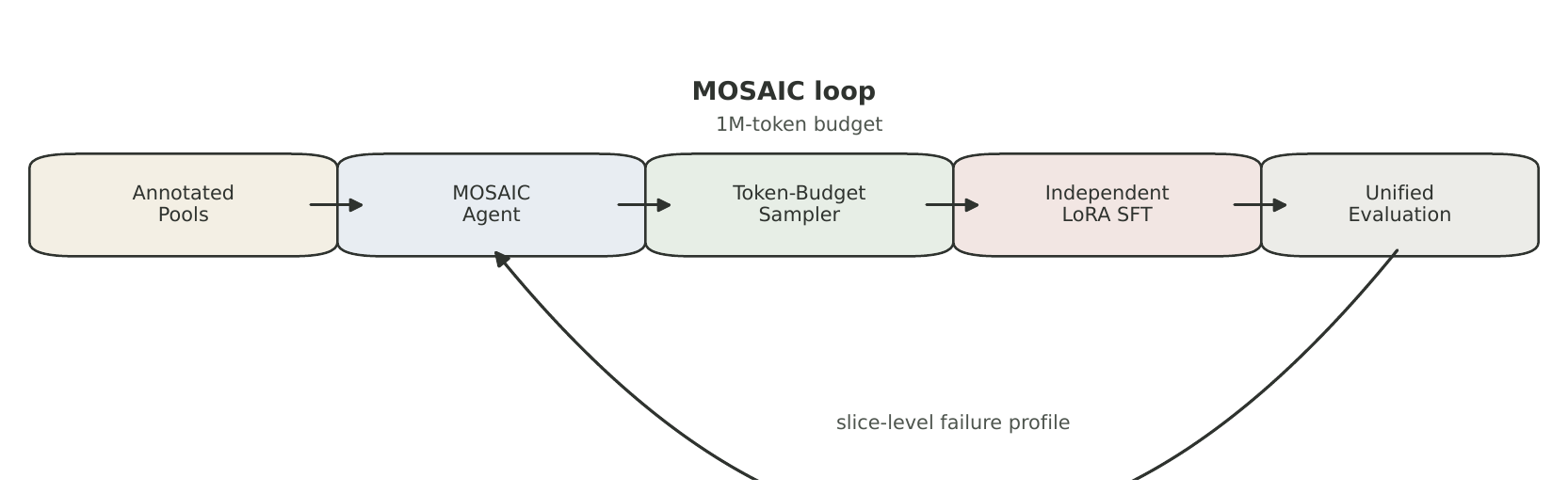}
    \caption{Overview of MOSAIC. The agent updates the next mixture from slice-level failure profiles under a fixed token budget.}
    \label{fig:framework}
\end{figure}

Figure~\ref{fig:framework} summarizes MOSAIC. A proposal agent reads historical results and the current failure profile, proposes a new data distribution, and triggers a fresh fine-tuning run. The same base model is reused for every round, which removes weight inheritance as a confounder. This design keeps the search target simple: the decision variable is the data mixture, not the model trajectory.

\subsection{Unified L1--L3 Evaluation Interface}

For each sample $i$, the prompt-side annotation emits three fields:
\begin{equation}
\mathrm{valid}_i \in \{0,1\}, \qquad s_i \in \mathcal{S}, \qquad w_i \in \mathbb{R}_{\ge 0},
\end{equation}
where $\mathrm{valid}_i$ indicates whether the sample participates in the main metric, $s_i$ is the slice label, and $w_i$ is the importance weight used for diagnostics and prioritization.

For each capability dimension $d \in \{\mathrm{Safe}, \mathrm{Benign}, \mathrm{IF}\}$, the response-side annotation emits a set of atomic L3 checks
\begin{equation}
a^{(d)}_{i,k} \in \{0,1\}, \qquad k = 1, \dots, K_d,
\end{equation}
which are aggregated into interpretable L2 states and then into a deterministic L1 score:
\begin{equation}
\mathrm{score}^{(d)}_i = F_d\!\left(\{a^{(d)}_{i,k}\}_{k=1}^{K_d}\right), \qquad \mathrm{score}^{(d)}_i \in [1,5].
\end{equation}

The crucial output is not the score alone, but a failure profile per slice:
\begin{equation}
\mathrm{FP}^{(d)}(s) =
\left(
\overline{\mathrm{score}}^{(d)}_s,
\mathrm{fail\_rate}^{(d)}_s,
\mathrm{breakdown}^{(d)}_s
\right).
\end{equation}
This tuple records how badly a slice performs, how often it fails, and which failure modes dominate.

\subsection{From Failure Profiles to Data Actions}

Let the searchable training buckets be $\mathcal{S}$. At iteration $t$, the system chooses a mixture vector
\begin{equation}
\mathbf{p}_t = (p_{t,1}, \dots, p_{t,|\mathcal{S}|}), \qquad
p_{t,j} \ge 0, \qquad
\sum_{j=1}^{|\mathcal{S}|} p_{t,j} = 1.
\end{equation}
Given a fixed token budget $B$, samples are drawn according to $\mathbf{p}_t$ until the budget is exhausted. Fine-tuning then starts from the same base model $\mathcal{M}_0$ and produces a round-specific model $\mathcal{M}_t$.

Each round returns a three-dimensional metric vector
\begin{equation}
\mathbf{m}_t = \left(M_t^{\mathrm{Safe}}, M_t^{\mathrm{Benign}}, M_t^{\mathrm{IF}}\right),
\end{equation}
plus slice-level failure profiles. The proposal agent observes the history
\[
\{(\mathbf{p}_{\tau}, \mathbf{m}_{\tau}, \mathrm{FP}_{\tau})\}_{\tau=1}^{t}
\]
and emits three action types for the next round:
\begin{enumerate}
    \item a dataset-level mixture over XGuard, OrBench, and IFEval data;
    \item bucket-level weights inside each dataset;
    \item focus criteria that prioritize particular L2/L3 failure patterns.
\end{enumerate}

Because the objectives conflict, we summarize results with the non-dominated set. For two metric vectors $\mathbf{m}_a$ and $\mathbf{m}_b$, we say $\mathbf{m}_a$ dominates $\mathbf{m}_b$ if
\begin{equation}
\mathbf{m}_a \succ \mathbf{m}_b
\iff
\left(\forall d,\; m_a^{(d)} \ge m_b^{(d)}\right)
\wedge
\left(\exists d,\; m_a^{(d)} > m_b^{(d)}\right).
\end{equation}

\subsection{Task-Specific Instantiations}

\paragraph{XGuard.}
XGuard evaluates multi-turn safety alignment against harmful or jailbreak-style conversations. Prompt-side slices are defined mainly by attack complexity, which is aggregated from pressure and concealment. Response-side checks capture explicit leakage channels such as step-by-step disclosure, code leakage, or bypass-style leakage, followed by the quality of the refusal and safe alternatives. This makes it possible to distinguish a merely non-compliant answer from a safe and helpful refusal.

\paragraph{OrBench.}
OrBench measures over-refusal on benign but boundary-adjacent requests. A sample is valid only if the prompt is fundamentally answerable through a safe path. Prompt-side slices combine behavior category and boundary proximity. Response-side annotations record refusal type, help level, and friction, so the framework can tell apart explicit refusal, evasive but nominally responsive answers, and genuinely helpful low-friction outputs.

\paragraph{IFEval.}
IFEval targets instruction following with verifiable constraints. Prompt-side slices reflect constraint family and complexity. Response-side checks are programmatic: they test hard constraints and overall satisfaction rate for format, length, inclusion, exclusion, and structural requirements. This makes the dimension deterministic and naturally auditable.

\section{Experimental Setup}

\subsection{Training Pools and Protocol}

The training pool combines three sources aligned with the three target capabilities. XGuard-Train provides multi-turn safety alignment data and expands from 30,695 raw conversations to 64,317 sliding-window training instances after preprocessing \cite{rahman2025xteaming}. For benign boundary calibration, we follow the OR-Bench construction paradigm but do not reuse its original training split directly; instead, we build a candidate pool from PKU-SafeRLHF prompts and coverage-guided sampling, yielding 8,226 training windows \cite{DBLP:conf/icml/CuiCSH25,DBLP:conf/iclr/DaiPSJXL0024}. For instruction following, we use the \texttt{allenai/tulu-3-sft-personas-instruction-following} dataset and retain 9,375 windows after filtering and segmentation.

All training runs use a 4,096-token sliding-window preprocessing strategy. The search is constrained by a fixed 1M-token budget per round and at most five fine-tuning iterations. Every round performs an independent LoRA run from the same Meta-Llama-3.1-8B-Instruct base model. Apart from the data distribution, all training hyperparameters are held constant.

\subsection{Evaluation and Hardware}

The internal search loop uses the unified L1--L3 interface on three evaluation sets: XGuard-style multi-turn attack traces, OrBench-style benign boundary prompts, and IFEval-style constraint prompts. In the main tables we report unweighted means for direct comparability across rounds; slice weights are used for prioritization and agent decisions rather than for headline reporting.

The experiments run on a single node with four NVIDIA A800-SXM4-80GB GPUs. Training is implemented in the PyTorch ecosystem and batched inference uses vLLM. Across the full five-round search, the end-to-end runtime is about 2.2 hours. On manually audited subsets, the structured XGuard and OrBench rubrics each achieve Cohen's $\kappa > 0.95$, which supports their use as reproducible diagnostic interfaces.

\section{Results}

\subsection{Closed-Loop Search under a Fixed Budget}

Table~\ref{tab:search_history} reports the full five-round search trajectory. The baseline model starts at $(2.76, 4.67, 3.43)$ on XGuard, OrBench, and IFEval. This profile exposes a clear imbalance: the base model is relatively permissive on benign boundary queries, moderately capable on instruction following, and weakest on safety alignment.

\begin{table}[t]
\centering
\caption{MOSAIC search history under a fixed 1M-token budget.}
\label{tab:search_history}
\small
\begin{tabular}{lcccl}
\toprule
\textbf{Round} & \textbf{XGuard} & \textbf{OrBench} & \textbf{IFEval} & \textbf{Dataset mixture (x/o/i)} \\
\midrule
Base & 2.7600 & \textbf{4.6667} & 3.4300 & no fine-tuning \\
0 & 3.2267 & 3.6433 & 3.5300 & 0.50 / 0.30 / 0.20 \\
1 & 3.3867 & 3.8033 & 3.5700 & 0.40 / 0.40 / 0.20 \\
2 & 4.4567 & 4.3300 & \textbf{3.7033} & 0.35 / 0.45 / 0.20 \\
3 & 3.9700 & 4.2967 & 3.5767 & 0.35 / 0.45 / 0.20 \\
4 & \textbf{4.6700} & 4.4067 & 3.6533 & 0.35 / 0.45 / 0.20 \\
\bottomrule
\end{tabular}
\end{table}

\begin{figure}[t]
    \centering
    \includegraphics[width=0.98\linewidth]{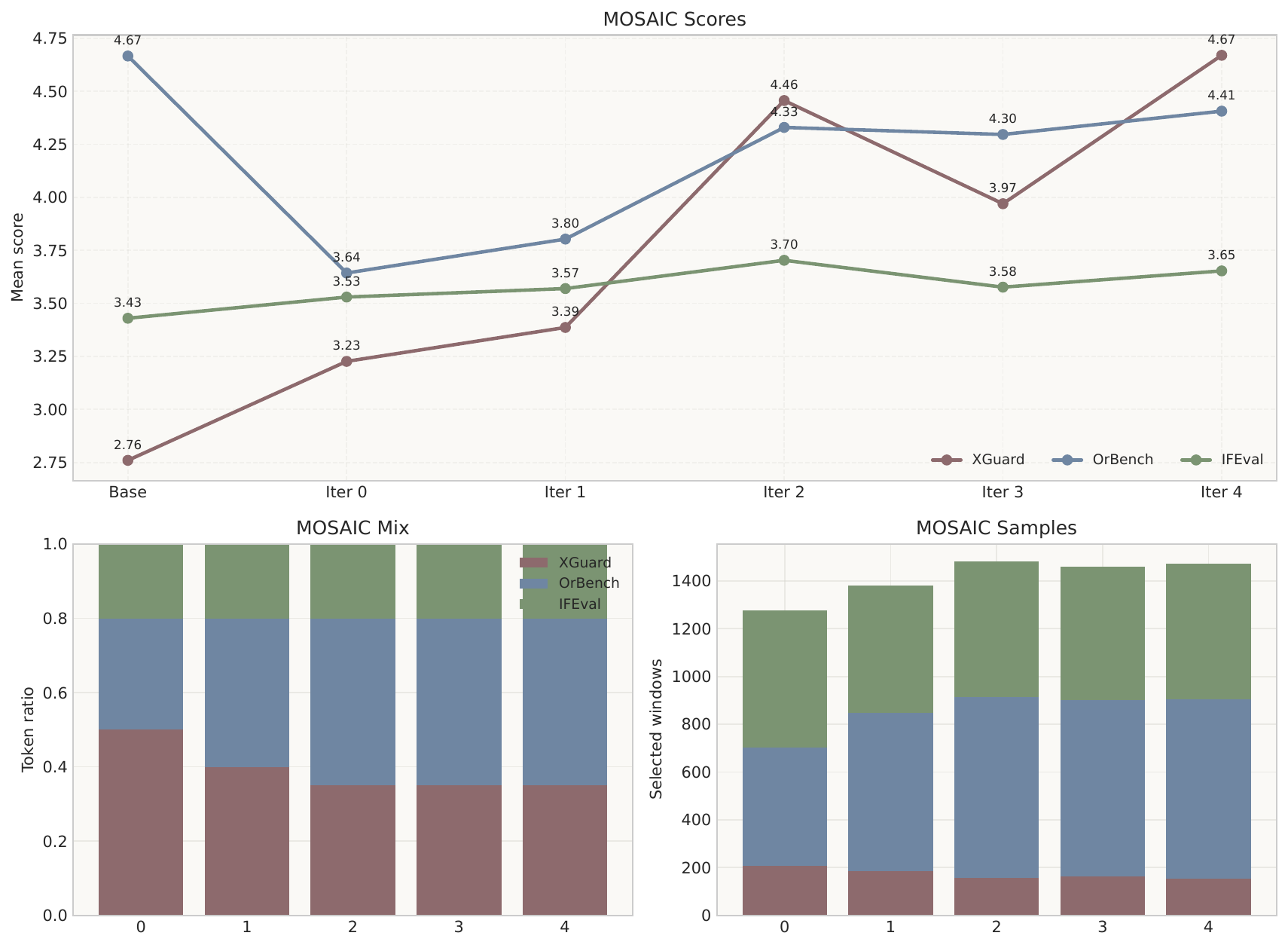}
    \caption{MOSAIC search trajectory under a fixed 1M-token budget. The proposal agent quickly converges on a stable macro mixture, while the number of selected windows still varies because the three datasets have different average lengths.}
    \label{fig:trajectory}
\end{figure}

Figure~\ref{fig:trajectory} shows the same process graphically. The first iteration exposes the central tension of the problem: pushing half of the budget into XGuard immediately improves safety from 2.76 to 3.23, but drops OrBench from 4.67 to 3.64. The next round partly repairs that damage by shifting budget toward OrBench. The decisive transition occurs at iteration 2, where the search converges to a macro mixture of $(0.35, 0.45, 0.20)$ and reaches a balanced point of $(4.46, 4.33, 3.70)$.

The best safety-oriented solution is iteration 4, which raises XGuard to 4.67 while keeping OrBench at 4.41 and IFEval at 3.65. Relative to the baseline, this is a gain of $+1.91$ on safety, a modest OrBench drop of only $0.26$, and a small IFEval improvement. The result shows that the trade-off is real but not strictly zero-sum.

\subsection{Pareto Frontier and Micro-Policy Effects}

The final non-dominated set contains three solutions, shown in Table~\ref{tab:pareto}. The baseline remains the best point for OrBench alone, iteration 2 is the most balanced three-objective solution and the best IFEval point, and iteration 4 is the best safety-oriented solution.

\begin{table}[t]
\centering
\caption{Non-dominated solutions found by MOSAIC.}
\label{tab:pareto}
\small
\begin{tabular}{lcccl}
\toprule
\textbf{Solution} & \textbf{XGuard} & \textbf{OrBench} & \textbf{IFEval} & \textbf{Interpretation} \\
\midrule
Base & 2.7600 & \textbf{4.6667} & 3.4300 & best no-tuning usability reference \\
Iter 2 & 4.4567 & 4.3300 & \textbf{3.7033} & most balanced three-objective point \\
Iter 4 & \textbf{4.6700} & 4.4067 & 3.6533 & strongest safety-oriented solution \\
\bottomrule
\end{tabular}
\end{table}

\begin{figure}[t]
    \centering
    \includegraphics[width=0.98\linewidth]{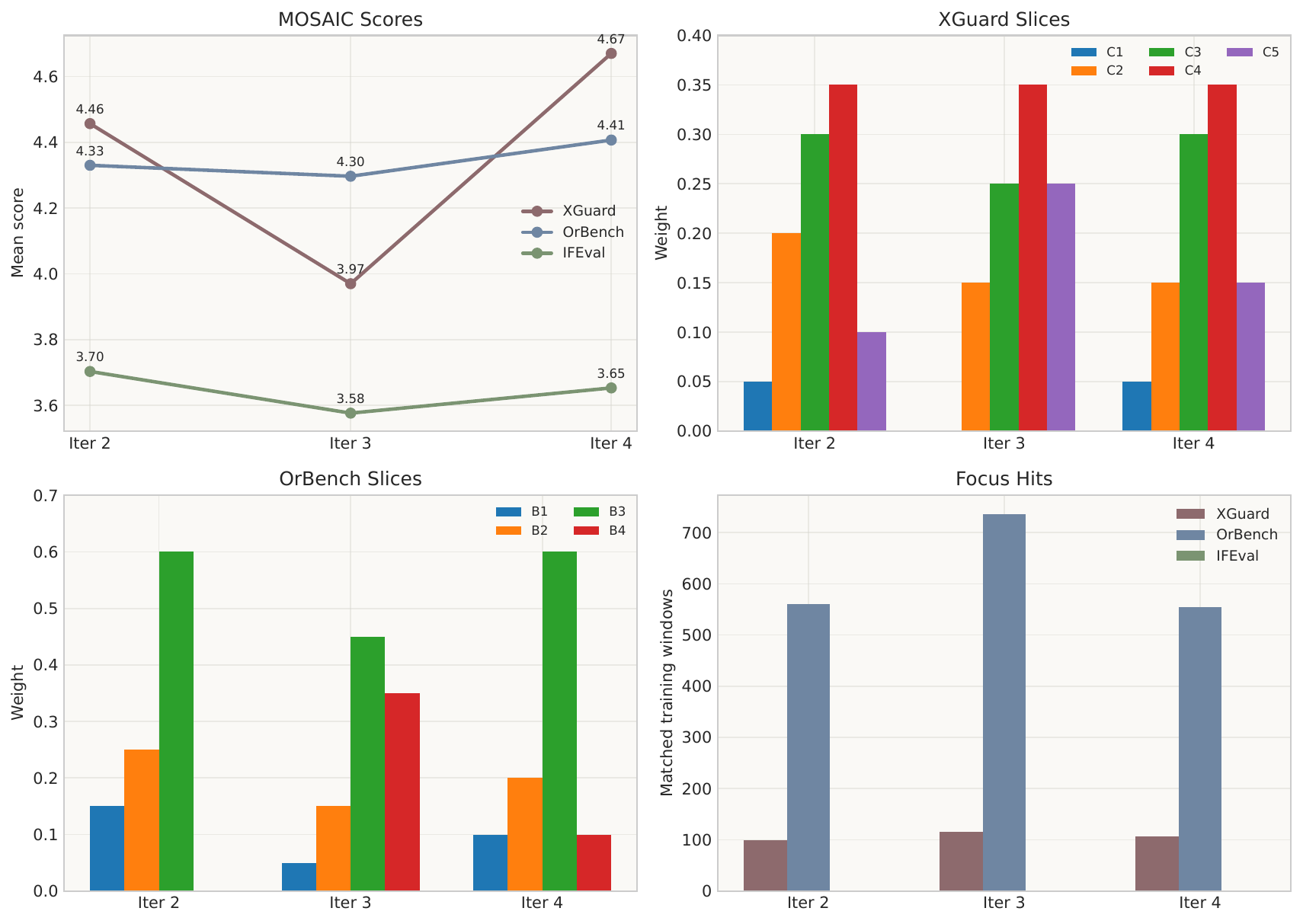}
    \caption{Controlled MOSAIC comparison for iterations 2--4. The macro mixture is identical, so the performance differences are attributable to bucket weights and focus criteria.}
    \label{fig:micro_policy}
\end{figure}

An important result is that dataset-level ratios stop changing after iteration 2, but outcomes do not. Iterations 2, 3, and 4 all use the same macro mixture $(0.35, 0.45, 0.20)$, yet their scores differ materially. Figure~\ref{fig:micro_policy} shows why: bucket weights and focus-hit counts shift even while the global budget remains fixed. The implication is that later-stage gains come from micro-level distribution control, not from further coarse reallocation between datasets.

Iteration 3 is especially revealing. Although it hits the largest number of OrBench-focused samples, its final scores regress to $(3.97, 4.30, 3.58)$. Iteration 4 recovers by restoring a small amount of low-complexity XGuard coverage, reducing the most extreme OrBench slice, and shrinking the focus scope. In other words, the framework does not reward blindly concentrating on the hardest cases; it rewards targeted rebalancing after the search has exposed where aggressive focus becomes counterproductive.

\begin{figure}[t]
    \centering
    \includegraphics[width=0.98\linewidth]{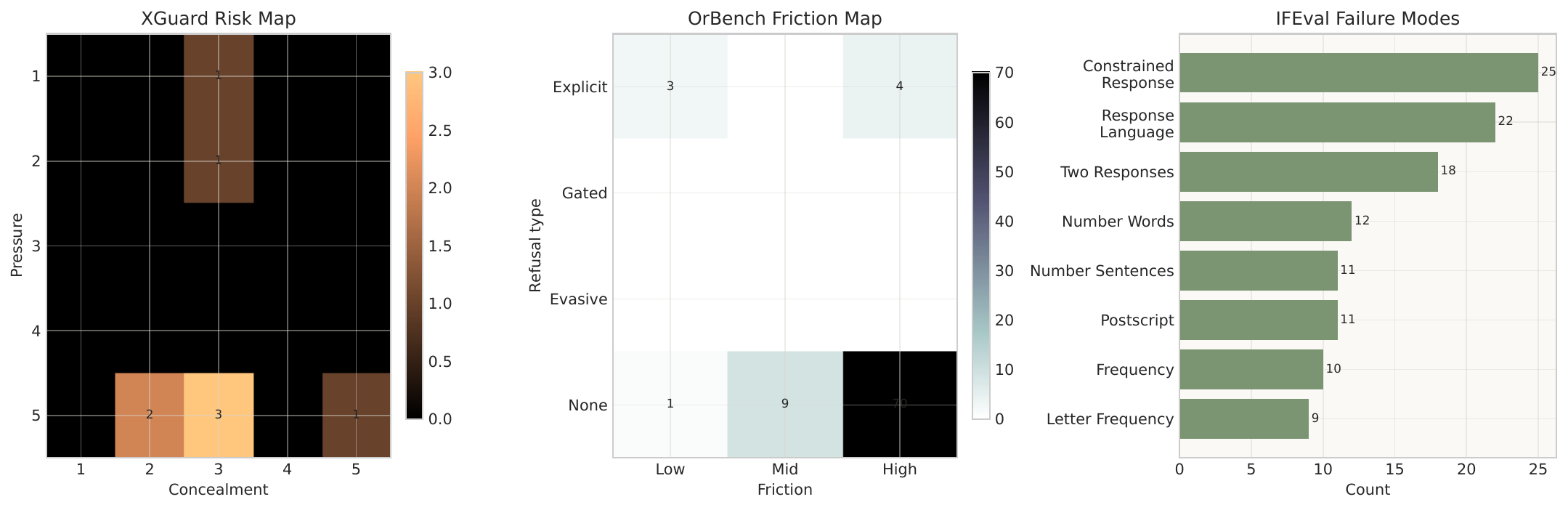}
    \caption{Diagnostic evidence from the final MOSAIC Pareto solution. The L1 score can be decomposed into concrete slice-level and atomic failure signals that directly inform data actions.}
    \label{fig:failure_breakdown}
\end{figure}

Figure~\ref{fig:failure_breakdown} illustrates how the three-layer interface makes those corrections possible. In the final safety-oriented solution, low-score XGuard samples cluster around high-pressure and high-concealment attacks, non-perfect OrBench outputs are often gated or high-friction rather than explicitly refusing, and IFEval errors collapse into a small set of frequent constraint families. These are exactly the signals consumed by the next-round sampler.

\subsection{External Validation against Random Static Mixing}

Internal search metrics are useful only if the resulting mixture generalizes. We therefore compare three models on independent external tests: the base model, a random static LoRA model trained on a random 1M-token mixture, and the final Pareto solution from iteration 4. Table~\ref{tab:external_summary} reports the results.

\begin{table}[t]
\centering
\caption{External validation of MOSAIC. Lower is better for attack success rate and over-refusal, higher is better for general capability.}
\label{tab:external_summary}
\small
\setlength{\tabcolsep}{4pt}
\begin{tabular}{l|ccc|cccc}
\toprule
& \multicolumn{3}{c|}{\textbf{Attack success rate (\%) $\downarrow$}} & \multicolumn{4}{c}{\textbf{Over-refusal (\%) $\downarrow$}} \\
\cmidrule(lr){2-4} \cmidrule(lr){5-8}
\textbf{Model} & \textbf{X-Teaming} & \textbf{ActorAttack} & \textbf{Crescendo} & \textbf{XSTest} & \textbf{OKTest} & \textbf{OR-Bench} & \textbf{PHTest} \\
\midrule
Base & 76.00 & 51.00 & 91.00 & 7.20 & \textbf{4.00} & 4.67 & 11.00 \\
Random LoRA & \textbf{1.00} & \textbf{21.50} & \textbf{17.00} & 18.00 & 26.67 & 17.67 & 36.33 \\
Pareto & 15.00 & 26.00 & 63.00 & \textbf{5.20} & 5.33 & \textbf{4.33} & \textbf{9.33} \\
\bottomrule
\end{tabular}

\vspace{0.7em}

\begin{tabular}{l|ccc}
\toprule
& \multicolumn{3}{c}{\textbf{General capability (\%) $\uparrow$}} \\
\cmidrule(lr){2-4}
\textbf{Model} & \textbf{MMLU} & \textbf{GSM8K} & \textbf{IFEval} \\
\midrule
Base & 68.48 & \textbf{81.73} & 74.68 \\
Random LoRA & 69.20 & 78.54 & 47.32 \\
Pareto & \textbf{69.26} & 81.20 & \textbf{78.19} \\
\bottomrule
\end{tabular}
\end{table}

The comparison is clear. Random static mixing suppresses attack success very aggressively, but it does so by becoming much more conservative overall. Over-refusal rises sharply on all four benign evaluation suites, and IFEval collapses from 74.68 to 47.32. The Pareto solution still reduces attack success substantially relative to the base model, but avoids the broad refusal inflation and preserves general capabilities.

\begin{figure}[t]
    \centering
    \includegraphics[width=0.98\linewidth]{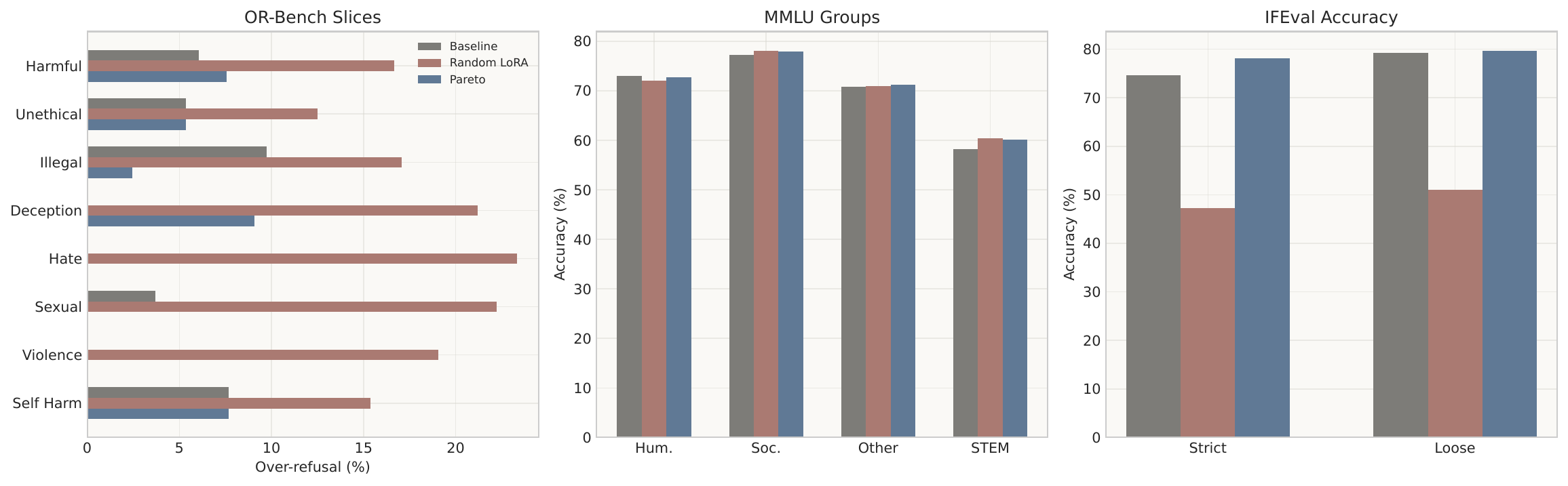}
    \caption{Granular external validation of the final MOSAIC solution. The Pareto model avoids the over-refusal spike and instruction-following collapse observed in random static mixing.}
    \label{fig:external_balance}
\end{figure}

Figure~\ref{fig:external_balance} provides a more granular view. The Pareto solution consistently lowers OR-Bench over-refusal across multiple behavior slices, remains stable across grouped MMLU categories, and preserves both strict and loose prompt-level IFEval performance. The gain is therefore not just a change in one summary number; it is a more balanced behavioral profile.

\section{Limitations}

This study has several limitations. First, the search covers only five iterations, so better Pareto solutions may exist beyond the explored trajectory. Second, the internal OrBench and IFEval search sets are derived from the same construction pipeline as their training pools, which makes them suitable for relative search decisions but not sufficient as standalone generalization evidence. Third, all experiments are run on a single base model, so the exact trade-off surface may differ for models with different initial alignment states. Fourth, we compare the proposal agent mainly against random static mixing; stronger search baselines such as rule-based proposals, random search, or Bayesian optimization remain for future work. Finally, some high-risk slices are sparse, which limits the statistical strength of the most fine-grained failure claims.

\section{Conclusion}

We presented MOSAIC, a closed-loop framework for data construction under fixed SFT budgets. The core idea is to replace opaque scalar feedback with a structured L1--L3 interface that exposes valid samples, slices, weights, atomic failures, and aggregated scores across three alignment objectives: safety alignment, over-refusal calibration, and instruction following. Because these signals map directly to executable data actions, MOSAIC can search over data mixtures rather than relying on static curation or ad hoc retuning.

Under a fixed 1M-token budget and independent fine-tuning in every round, the framework finds non-dominated solutions that improve safety substantially while preserving usability and instruction following far better than random static mixing. The broader implication is practical: when evaluation outputs are made interpretable enough to support slice-level diagnosis, they can serve as auditable control signals for real data allocation decisions.

\bibliographystyle{unsrt}
\bibliography{references}

\end{document}